\documentclass[useAMS,usenatbib,usegraphicx]{mn2e}

\title[ACS photometry of ESO 121-SC03]{Photometry of Magellanic Cloud clusters 
with the Advanced Camera for Surveys - II. The unique LMC cluster ESO 121-SC03}
\author[A.~D.~Mackey et al.]{A.~D.~Mackey,$^{1}$\thanks{E-mail:
dmackey@ast.cam.ac.uk} M.~J.~Payne$^{1}$ and G.~F.~Gilmore$^{1}$\\
$^{1}$Institute of Astronomy, University of Cambridge, Madingley Road, 
Cambridge CB3 0HA}
\begin{document}

\date{Accepted 2006 March 17. Received 2006 March 17; in original form 2006 March 03.}

\pagerange{\pageref{firstpage}--\pageref{lastpage}} \pubyear{2006}

\maketitle

\label{firstpage}

\begin{abstract}
We present the results of photometric measurements from images of the LMC cluster
ESO 121-SC03 taken with the Advanced Camera for Surveys on the {\it Hubble Space
Telescope}. Our resulting colour-magnitude diagram (CMD) reaches $3$ magnitudes below the
main-sequence turn-off, and represents by far the deepest observation of this
cluster to date. We also present similar photometry from ACS imaging of the accreted 
Sagittarius dSph cluster Palomar 12, used in this work as a comparison cluster.
From analysis of its CMD, we obtain estimates
for the metallicity and reddening of ESO 121-SC03: $[$Fe$/$H$] = -0.97 \pm 0.10$ 
and $E(V-I) = 0.04 \pm 0.02$, in excellent agreement with previous studies. The 
observed horizontal branch level in ESO 121-SC03 suggests this cluster may lie $20$
per cent closer to us than does the centre of the LMC. ESO 121-SC03 also possesses
a significant population of blue stragglers, which we briefly discuss.
Our new photometry allows us
to undertake a detailed study of the age of ESO 121-SC03 relative to Pal. 12 and the 
Galactic globular cluster 47 Tuc. We employ both vertical and horizontal differential 
indicators on the CMD, calibrated against isochrones from the Victoria-Regina stellar 
models. These models allow us to account for the different $\alpha$-element abundances
in Pal. 12 and 47 Tuc, as well as the unknown run of $\alpha$-elements in ESO 121-SC03.
Taking a straight error-weighted mean of our set of age measurements yields ESO 121-SC03 
to be $73 \pm 4$ per cent the age of 47 Tuc, and $91 \pm 5$ per cent the age of Pal. 
12. Pal. 12 is $79 \pm 6$ per cent as old as 47 Tuc, consistent with previous work.
Our result corresponds to an absolute age for ESO 121-SC03 in the range $8.3-9.8$ Gyr,
depending on the age assumed for 47 Tuc, therefore confirming ESO 121-SC03 as the only
known cluster to lie squarely within the LMC age gap. We briefly discuss a suggestion
from earlier work that ESO 121-SC03 may have been accreted into the LMC system.
\end{abstract}


\section{Introduction}
\label{s:intro}
The Large Magellanic Cloud (LMC) contains an extensive population of massive star 
clusters, with ages ranging from the very newly formed (e.g., R136 with age $\sim 3-4$ 
Myr) to those with ages comparable to the Galactic globular clusters (e.g., Hodge 11).
These objects are extremely useful not only for studies of cluster development,
but also for tracing the formation history of the LMC and its related kinematic, 
structural, and chemical evolution.

Of particular interest in this respect is the distribution of star cluster ages. A 
number of authors have explicitly demonstrated, using high quality colour-magnitude
diagrams, that the oldest globular clusters in the LMC are coeval with the oldest 
Galactic globular clusters, to within $\sim 1$ Gyr \citep[e.g.,][]{johnson:99,olsen,mackey}.
With the recent addition of NGC 1928 and 1939, the census of objects constituting this
group of ancient clusters numbers $15$ \citep{mackey}. There also exists a very 
substantial population of young and intermediate age clusters; however, examination of
the age distribution reveals an almost complete dearth of clusters older than $\sim 3$ Gyr
but younger than the ancient ensemble \citep[e.g.,][]{rich}. This curious distribution
constitutes the well known LMC cluster age-gap. Recent research aimed at explaining the
existence of the age-gap has focused on the impact of the periodic strong tidal
interactions between the LMC and its companion the Small Magellanic Cloud (SMC)
\citep{bekki}.

Only one cluster is suspected to lie within the age gap. This is the remote northern
LMC member ESO 121-SC03, which lies at a projected angular separation of $\sim 10\degr$ 
from the LMC centre. ESO 121-SC03 was first studied in detail by \citet{mateo}, who 
used $BV$ photometry to estimate an age in the range $8-10$ Gyr. Subsequent work has 
not altered this conclusion \citep[e.g.,][]{sarajedini:95,geisler,bica}. Given its 
apparently unique age, and interesting spatial location, ESO 121-SC03 clearly
constitutes an object worthy of detailed study. We were therefore surprised to discover
that it apparently has not been the subject of deep high-resolution imaging:
we could only locate two independent published CMDs in the literature -- the $BV$
photometry of \citet{mateo}, and the Washington photometry of \citet{geisler}.
Neither of these studies reaches more than $\sim 1$ mag below the main-sequence turn-off
(MSTO), and indeed in both, the turn-off is only measured at low signal-to-noise.

In this paper we take advantage of the existence of observations of ESO 121-SC03 taken
with the Advanced Camera for Surveys (ACS) on {\it HST} as part of our recent snapshot
survey of LMC and SMC star clusters (program 9891), to make deep, high quality
photometric measurements of this cluster. Our resulting $VI$ CMD reaches more than
$3$ magnitudes below the main-sequence turn-off, representing by far the deepest 
observations of ESO 121-SC03 to date. We use these data to conduct a detailed study
of the age of ESO 121-SC03 relative to two comparison clusters -- the young accreted
Sagittarius dSph globular cluster Palomar 12 (also imaged with ACS in program 9891), 
and the old Galactic globular cluster 47 Tuc. We find that ESO 121-SC03 is significantly 
younger than 47 Tuc, but very similar in age to Pal. 12, confirming the earlier 
conclusion of \citet{mateo} and \citet{geisler} that ESO 121-SC03 does indeed lie 
squarely within the LMC cluster age-gap.

\section{Observations and Data Reduction}
\label{s:obsred}
The observations were taken during {\em HST} Cycle 12 using the ACS Wide Field 
Channel (WFC). As snapshot targets, the clusters were observed for only one orbit each. 
This allowed two exposures to be taken per cluster -- one through the F555W filter and 
one through the F814W filter. Details of the individual exposures are listed in 
Table \ref{t:observations}. 

The ACS WFC consists of a mosaic of two $2048 \times 4096$ SITe CCDs with a scale 
of $\sim 0.05$ arcsec per pixel, and separated by a gap of $\sim 50$ pixels. Each 
image therefore covers a field of view (FOV) of approximately $202 \times 202$ 
arcseconds. The clusters were centred at the reference point WFC1, located on chip 1 
at position $(2072\,,\,1024)$. This allowed any given cluster to be observed up to a 
radius $r\sim 150 \arcsec$ from its centre, while also ensuring the cluster core did 
not fall near the inter-chip gap.

Our observations were made with the ACS/WFC GAIN parameter set to $2$. This allows 
the full well depth to be sampled (as opposed to only $\sim 75$ per cent of the full 
well depth for GAIN $=1$) with only a modest increase in read noise ($\sim 0.3$ 
e$^{-}$ extra rms), thus increasing the dynamic range of the observations by
greater than $0.3$ mag over that obtained with the default gain. In addition, we
offset the second image of each cluster by $2$ pixels in both the $x$ and $y$ 
directions, to help facilitate the removal of hot pixels and cosmic rays. With only 
two images per cluster, through different filters, it is not possible to completely
eliminate the inter-chip gap using such an offset.

\begin{table*}
\begin{minipage}{164mm}
\caption{ACS/WFC observations of ESO 121-SC03 and Pal. 12 ({\em HST} program 9891).}
\begin{tabular}{@{}lcccccccc}
\hline \hline
Cluster & \hspace{10mm} & RA & Dec. & \hspace{10mm} & Filter & Image & Exposure & Date \\
 & & (J2000.0) & (J2000.0) & & & Name & Time (s) & \\
\hline
ESO 121-SC03 & & $06^{{\rm h}}\,\,02^{{\rm m}}\,\,01.36^{{\rm s}}$ & $-60\degr\,\,31\arcmin\,\,22.6\arcsec$ & & F555W & j8ne79sdq & $330$ & $07/10/2003$ \\
 & & & & & F814W & j8ne79smq & $200$ & $07/10/2003$ \\
Palomar 12 & & $21^{{\rm h}}\,\,46^{{\rm m}}\,\,39.18^{{\rm s}}$ & $-21\degr\,\,15\arcmin\,\,06.8\arcsec$ & & F555W & j8ne80koq & $53$ & $26/07/2003$ \\
 & & & & & F814W & j8ne80kqq & $32$ & $26/07/2003$ \\
\hline
\label{t:observations}
\end{tabular}
\end{minipage}
\end{table*}

During retrieval from the STScI archive, all images were passed through the standard 
ACS/WFC reduction pipeline. This process includes bias and dark subtractions,
flatfield division, masking of known bad pixels and columns, and the calculation of 
photometry header keywords. In the final stage of the pipeline, the {\sc pyraf}
{\sc multidrizzle} software is used to correct the (significant) geometric distortion 
present in WFC images. The products obtained from the STScI archive are hence fully 
calibrated and distortion-corrected drizzled images, in units of counts per second.

For both clusters, we performed photometry on the F555W and F814W images individually, 
using the {\sc daophot} software in {\sc iraf}. First the {\sc daofind} task was 
used with a detection threshold of $4\sigma$ above background to locate all the 
brightness peaks in each image. The two output lists were matched against each other 
to find objects falling at identical positions in the two frames. Objects detected in 
one image but with no matching counterpart in the other image were discarded.
Because of the significant geometric distortion present in ACS/WFC images, a 
$2\times 2$ pixel offset in telescope position does not correspond to a uniform
$2\times 2$ pixel offset between the two drizzled frames; in fact the offset is
position dependent. To match between the two lists, we therefore transformed all 
measured positions into the coordinate systems of the original distorted flat-fielded 
images and matched between these new lists. For this transformation we employed
the {\sc pyraf} task {\sc tran}. 

Next, we used the {\sc phot} task to perform aperture photometry on all objects in
the cross-matched lists, on each of the two images. For this we used an aperture
radius of $4$ pixels. This radius was selected after some experimentation, which 
demonstrated it to give the narrowest sequences on the two cluster CMDs. We also
performed measurements using the recently developed {\sc dolphot} software
(similar to {\sc HSTphot}, described by \citet{hstphot}); however we found that 
our aperture photometry pipeline consistently returned narrower CMD sequences for 
our two clusters. We therefore selected the aperture photometry as our preferred 
measurements for the present study. It is important to note both ESO 121-SC03 and 
Pal. 12 are very sparse clusters, so crowding is not an issue in the ACS/WFC 
images, and aperture photometry is a perfectly acceptable technique. Comparing the 
offsets between our aperture photometry (after cleaning and the application of the 
corrections and zero-points described below) and the {\sc dolphot} measurements 
revealed mean magnitude-independent offsets of $\la 0.01$ mag in both F555W and F814W. 
Hence we are confident in the accuracy of our photometric measurements.

We completed the photometric calibration using the results of \citet{sirianni}.
Magnitudes were calculated in the ACS/WFC VEGAMAG system, defined as the system
in which Vega has magnitude $0$ in all filters. From \citet{sirianni}, the 
relevant zero-points for this system are $25.724$ in F555W and $25.501$ in F814W.

Like all previous {\em HST} CCD instruments, the ACS/WFC chips are suffering from 
degradation of their charge transfer efficiency (CTE) due to radiation damage. 
This can cause position dependent errors in photometric measurements of up to
$\sim 10$ per cent under certain circumstances (see \citet{sirianni} and \citet{cte}
for detailed discussions). To correct for the charge loss, we employed the 
calibration of \citet{cte}, who provide a time-dependent parametrization of the 
necessary correction due to parallel ($y$-direction) transfers:
\begin{equation}
\Delta_Y = 10^A \times s^B \times f^C \times \frac{Y}{2048} \times \frac{(MJD - 52333)}{365}\,\,\,{\rm mag}
\label{e:cte}
\end{equation}
for any given object. In this equation, the object's sky ($s$) and flux ($f$) values 
are in counts; $Y$ represents the number of parallel transfers (so if the object 
has position $(x\,,\,y)$ on either of the two ACS chips, then $Y=y$ for objects on
chip 2, and $Y=2048-y$ for objects on chip 1); and MJD is the Modified Julian Date 
of the observation being corrected. The parameters $A$, $B$, and $C$ are tabulated
by \citet{cte} for three different apertures ($3$, $5$, and $7$ pixels). As our
$4$ pixel aperture has not been calibrated, we used the average of the corrections 
necessary for $3$ and $5$ pixel apertures. In the presently available calibration
there is no requirement for corrections due to serial ($x$-direction) transfers.

Our final step was to apply aperture corrections to the photometry. For this, we
followed the process described by \citet{sirianni}. For the F555W measurements,
we selected a sub-sample of suitable stars (see below) and used the {\sc phot}
task to measure photometry with a $0.5\arcsec$ aperture ($10$ WFC pixels). This 
allowed the correction from our $4$ pixel aperture to $0.5\arcsec$ to be calculated.
We then applied the correction from a $0.5\arcsec$ aperture to an infinite aperture
listed by \citet{sirianni} (AC05 in their terminology, where AC05\ $= 0.092$ mag
in F555W). The procedure for F814W was more complicated, as in the near-IR ACS filters
the aperture correction is a function of colour for red stars. This is due to the
scattering of red light in the CCDs, which results in a broadened PSF and 
long-wavelength halo \citep{sirianni}. We therefore used the {\sc calcphot} routine
in {\sc synphot} and the BPGS stellar atlas available in {\sc stsdas} to determine 
the relationship between the instrumental colour F555W-F814W of any given star and
its effective wavelength, as described by Sirianni et al. Here, the effective 
wavelength is a parameter which represents the mean wavelength of detected photons:
\begin{equation}
\lambda_{{\rm eff}} = \frac{\int f_{\lambda}(\lambda) P(\lambda) \lambda^2 d\lambda}{\int f_{\lambda}(\lambda) P(\lambda) \lambda d\lambda}\,,
\label{e:efflambda}
\end{equation}
where $P(\lambda)$  is the passband transmission curve (F814W in our case), and
$f_{\lambda}(\lambda)$ is the flux distribution of the object. Having defined this
relationship, we estimated the effective wavelength of each detected star from
the measured instrumental colour. A suitable subset of stars with blue effective
wavelength\footnote{Specifically, stars with $\lambda_{{\rm eff}} \le 8200 {\rm \AA}$.
In practice, the bluest stars have $\lambda_{{\rm eff}} \sim 8000 {\rm \AA}$ in F814W.} 
were used to calculate the aperture correction to $0.5\arcsec$, just as with the F555W 
measurements. This value, along with AC05 (equal to $0.087$ in F814W) was applied 
for blue stars. For all redder stars (i.e., with 
$\lambda_{{\rm eff}} > 8200 {\rm \AA}$), wavelength-dependent aperture corrections 
were applied directly from Table 6 of \citet{sirianni}. 

\begin{figure*}
\begin{minipage}{175mm}
\begin{center}
\includegraphics[width=87mm]{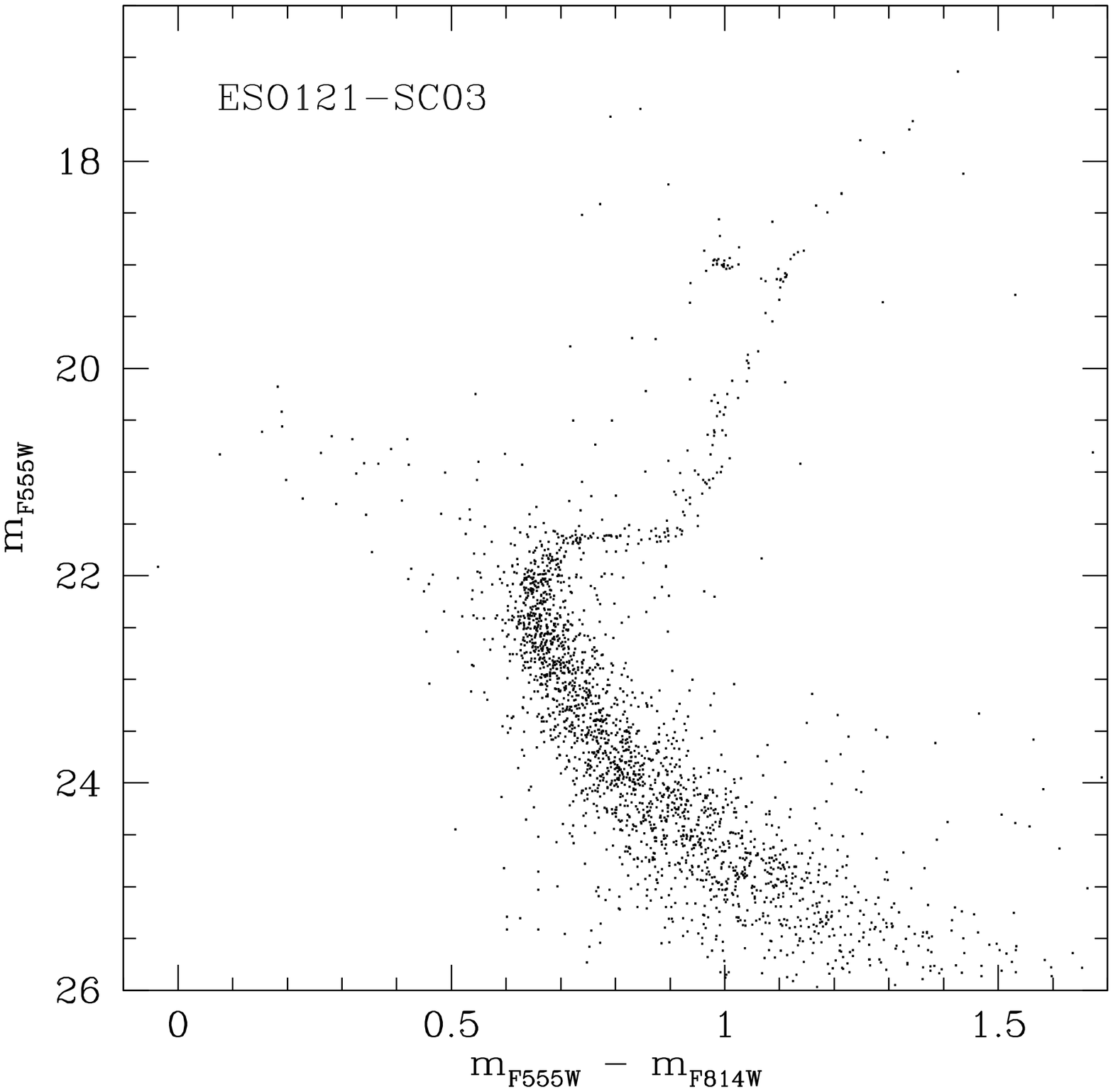}
\hspace{-2mm}
\includegraphics[width=87mm]{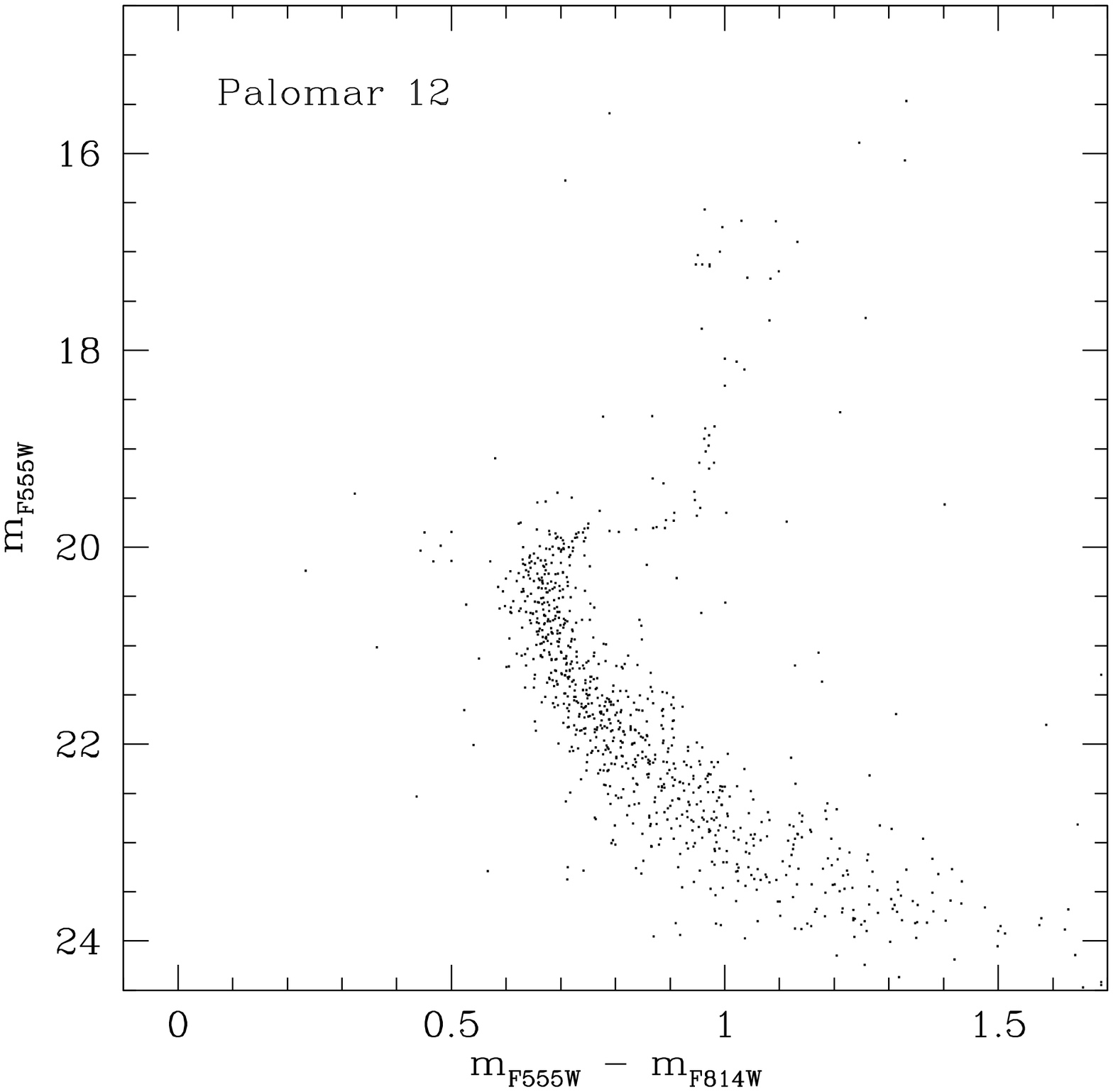}
\caption{Cleaned colour-magnitude diagrams for ESO 121-SC03 and Palomar 12. Measurements 
are plotted in the VEGAMAG magnitude system (see text). The CMD for ESO 121-SC03 
contains $2\,386$ detections, while that for Palomar 12 contains $944$ detections.}
\label{f:cmdvegamag}
\end{center}
\end{minipage}
\end{figure*}

In selecting suitable aperture stars, we applied the following criteria: they must 
be bright stars but significantly below saturation; they must not have unusual shape 
characteristics (see Section \ref{ss:colourmag}); they must have no neighbouring stars, 
bad pixels, cosmic rays, or image edges within a radius of $1\arcsec$; and they must 
not lie in an area of unusually high background (e.g., near a very bright star). In 
addition, for the F814W selection, they must have 
$\lambda_{{\rm eff}} \le 8200 {\rm \AA}$. 
For ESO 121-SC03, these defined a sample of $57$ stars in F555W and $90$ stars in
F814W; while for Palomar 12 they defined samples of $231$ and $210$ stars respectively.
The mean aperture correction to $0.5\arcsec$ was calculated for each image using
a $3\sigma$ clipping algorithm. Standard errors in the calculated corrections were 
typically $\sim 0.01$ mag.

\section{Results}
\label{s:results}
\subsection{Colour-magnitude diagrams}
\label{ss:colourmag}
We present the final colour-magnitude diagrams (CMDs) for ESO 121-SC03 and Pal. 12 in 
Fig. \ref{f:cmdvegamag}. Both stellar samples have been cleaned using the
PSF shape parameters (e.g., sharpness, roundness) obtained during the photometric
measurement process. This filtering is necessary to help remove cosmic rays and other
spurious or non-stellar detections. Because of the nature of the snapshot observations, 
only one image was taken in each filter; hence more cosmic rays propagate through the
photometry pipeline than in the usual scenario where multiple images per passband
are median filtered prior to the photometric measurements in order to remove spurious 
objects. 

To apply the quality filter we used the three shape characteristics from {\sc daofind}: 
one sharpness and two roundness parameters. The sharpness is a measurement of how high
the peak of a detection is relative to a best-fitting Gaussian, while the two
roundness parameters are measurements of how circular an object is. Clean stellar
detections should have sharpness $\sim 0.75$ and roundness $\sim 0$. For both photometry
lists we produced histograms of the three relevant parameters to determine suitable
clipping limits. Because both targets are sparse clusters, we set rather lenient limits
in order to maintain clear CMD sequences. For both, we only retained objects with
$0.6 \le {\rm sharpness} \le 0.9$ and objects with $-0.5 \le {\rm roundness} \le 0.5$
through each passband. This typically removed about $\sim 30$ per cent of detections,
predominantly at the faintest magnitudes.

We did not apply any filtering to remove field star contamination, as both clusters are
set against sparse fields -- ESO 121-SC03 is projected $\sim 10 \degr$ from the optical
centre of the LMC, while Pal. 12 lies at a Galactic latitude of almost $-50 \degr$. The
observed field of view is not sufficiently large to make an accurate estimate of the
surface density of field stars, in order to make an accurate subtraction.

The CMDs for both clusters show stubby red horizontal branches, consistent with previous
studies \citep[e.g.,][]{mateo,rosenberg}. Both CMDs also show populations of blue straggler
candidates. It is already known that Pal. 12 possesses such objects 
\citep[e.g.,][]{rosenberg}; however we know of no discussion of blue stragglers in
ESO 121-SC03. We discuss the blue straggler population of this cluster in more detail 
below (Section \ref{ss:bss}).

\begin{figure*}
\begin{minipage}{175mm}
\begin{center}
\includegraphics[width=87mm]{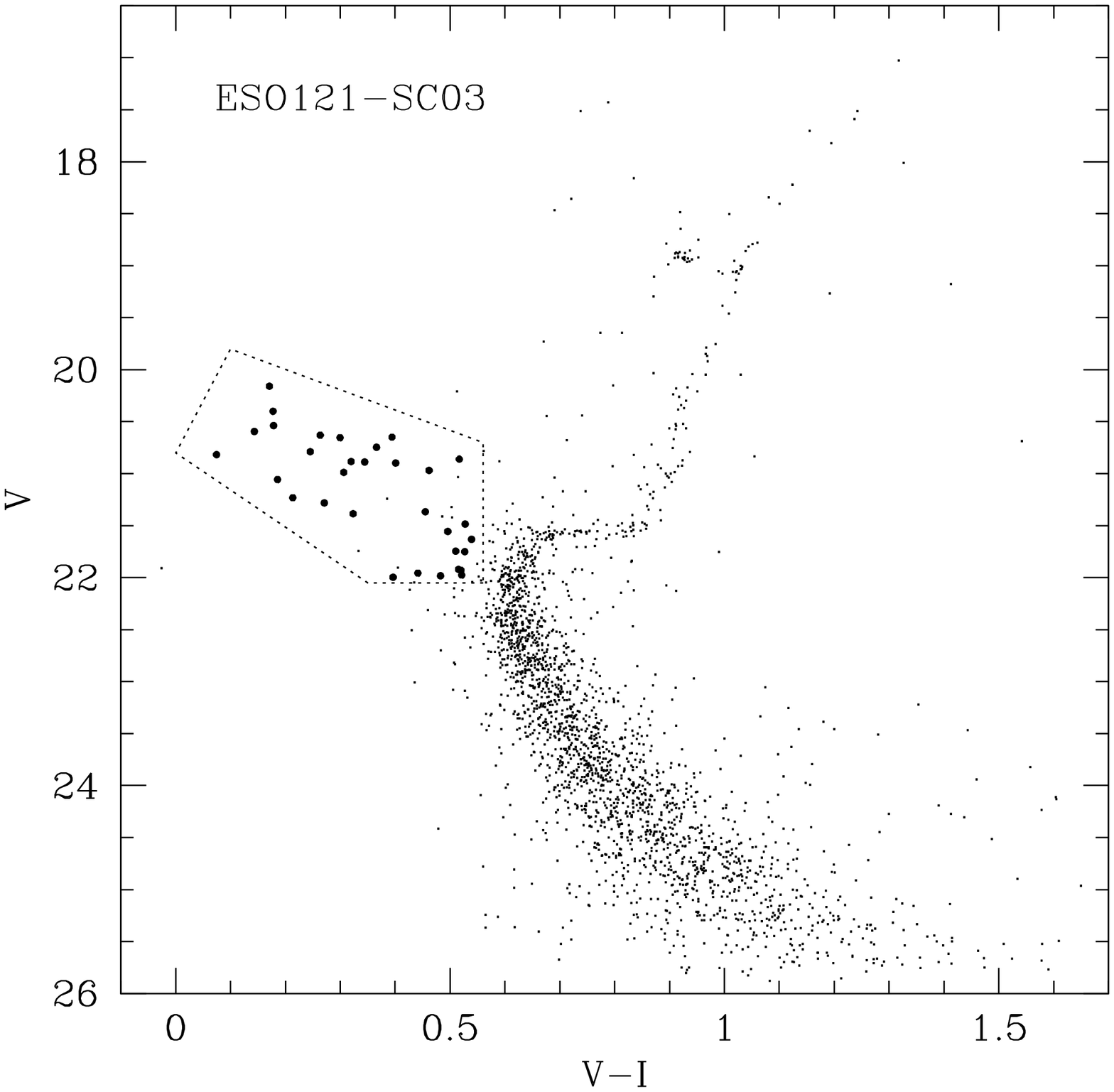}
\hspace{-2mm}
\includegraphics[width=87mm]{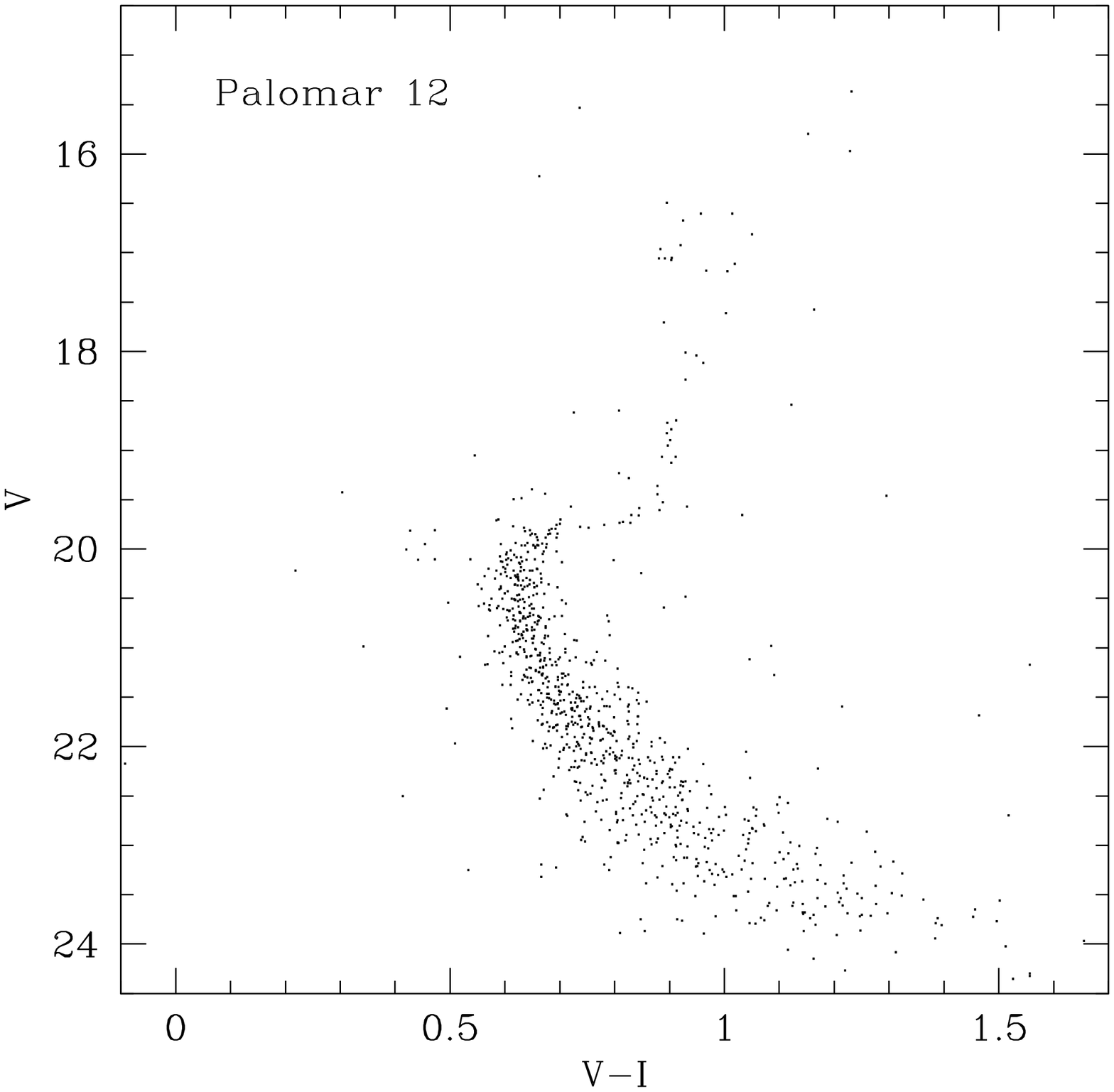}
\caption{Cleaned colour-magnitude diagrams for ESO 121-SC03 and Palomar 12, where the
photometry has been transformed from the VEGAMAG system to Johnson-Cousins $V$ and $I$
according to the prescription of \citet{sirianni} (see also text). On the CMD for
ESO 121-SC03 we have marked a region defining the blue straggler population for this
cluster. As described in the text, $40$ stars lie in this region, $32$ of which are very 
clean stars on the ACS images (bold points).}
\label{f:cmdvimag}
\end{center}
\end{minipage}
\end{figure*}

Our CMD for ESO 121-SC03 reaches $\sim 3$ mag below the main-sequence turn-off,
as does that for Pal. 12. This is the highest quality published CMD for ESO 121-SC03,
and offers the opportunity for an accurate relative age measurement. While we would
prefer to work in the native ACS photometric system to perform such a measurement,
we could not locate suitably sampled stellar isochrones (which we require in order to 
establish the relative age calibration) calculated in this system. We therefore used
the relations derived by \citet{sirianni} to transform our photometry into standard 
Johnson-Cousins $V$ and $I$ magnitudes. These relations take the form of an
iterative transformation:
\begin{equation}
{\rm TMAG} = {\rm SMAG} + c_0 + (c_1\times{\rm TCOL}) + (c_2\times{\rm TCOL}^2)
\label{e:transform}
\end{equation}
where the target magnitude and colour are TMAG and TCOL respectively, the source 
magnitude is SMAG, and $c_0$, $c_1$ and $c_2$ are the transformation coefficients
listed in Table 22 of \citet{sirianni}. To apply the transformation for a given
star, we first subtract the VEGAMAG zeropoints to obtain SMAG in F555W and F814W.
This also yields the source colour F555W-F814W, which we adopt as a crude first estimate
of TCOL. Using the appropriate coefficients in Eq. \ref{e:transform} then provides
first estimates of TMAG in $V$ and $I$. These in turn give an improved estimate
of TCOL, and the procedure is iterated until convergence. 

Fig. \ref{f:cmdvimag} presents the CMDs for ESO 121-SC03 and Pal. 12 in the standard
Johnson-Cousins photometric system. By comparing these to the CMDs in Fig. 
\ref{f:cmdvegamag} it is clear that the transformations have not drastically 
reshaped the colour-magnitude plane. This observation, along with the fact that
Sirianni et al's transformations were derived using a Galactic globular 
cluster\footnote{i.e., a stellar population not very different to those examined
here -- although admittedly the cluster in question (NGC 2419) is $\sim 1$ dex more
metal poor than either ESO 121-SC03 or Pal. 12.} leads us to have confidence that
we have not introduced damaging systematic errors into our photometry. We verify
this more explicitly in Section \ref{ss:metallicity}, below.

\subsection{Photometric metallicities and reddenings}
\label{ss:metallicity}
Although a number of metallicity estimates already exist for both ESO 121-SC03 and 
Pal. 12, our new observations provide an opportunity to add additional measurements.
Making these also allows us to check the consistency of our transformed photometry.
We employ the calibration of \citet{sarajedini}, who used high quality observations 
of six Galactic globular clusters to derive a method of simultaneously estimating
the reddening and metallicity of a cluster from its $(V,\,V-I)$ CMD. His method relies 
on the measurement of the level of the horizontal branch (HB) and the colour of the
red-giant branch (RGB) at the HB level, together with some parametrization of the RGB 
fiducial above the HB level. The calibration then relates the height of the RGB
above the HB at intrinsic colour $(V-I)_0 = 1.2$ to the metallicity (this parameter
is labelled $\Delta V_{1.2}$), and the measured colour of the RGB at $V_{{\rm HB}}$
(labelled $(V-I)_g$) to the reddening.

\begin{figure}
\includegraphics[width=0.5\textwidth]{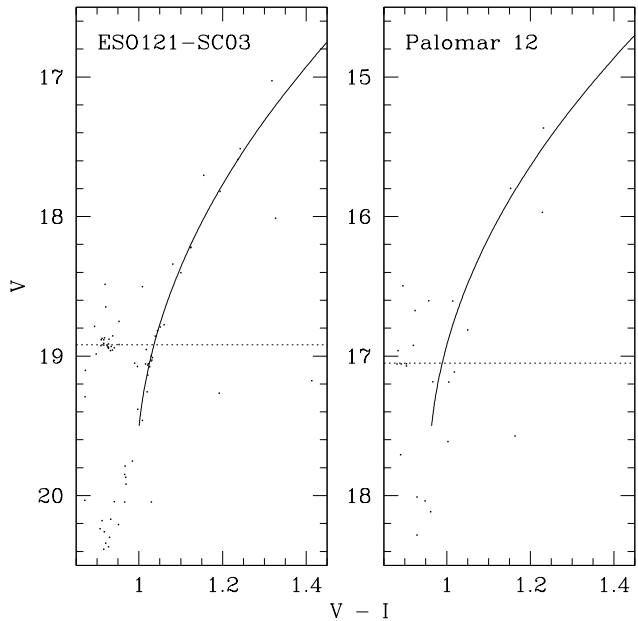}
\caption{Quadratic fits to the two cluster RGBs, used for the reddening and metallicity
estimates. The dotted lines indicate the HB levels. Note that the brightest star in
the ESO 121-SC03 red-giant branch (at $V\sim 17$) was not included in the fit for this
cluster as it is saturated in the F814W image and hence has large colour uncertainty.}
\label{f:rgbfits}
\end{figure}

\begin{table*}
\begin{minipage}{159mm}
\caption{Results of the simultaneous determination of cluster reddenings and metallicities.}
\begin{tabular}{@{}lccccccccc}
\hline \hline
Cluster & \hspace{3mm} & $V_{{\rm HB}}$ & $a_0$ & $a_1$ & $a_2$ & \hspace{3mm} & $\Delta V_{1.2}$ & $[$Fe$/$H$]$ & $E(V-I)$ \\
\hline
ESO 121-SC03 & & $18.90 \pm 0.03$ & $19.60150$ & $-1.87403$ & $0.04719$ & & $1.31 \pm 0.10$ & $-0.97 \pm 0.10$ & $0.04 \pm 0.02$ \\
Palomar 12 & & $17.05 \pm 0.03$ & $16.84290$ & $-1.78086$ & $0.04991$ & & $1.40 \pm 0.10$ & $-1.05 \pm 0.10$ & $0.00 \pm 0.02$ \\
\hline
\label{t:metred}
\end{tabular}
\end{minipage}
\end{table*}

We chose to adopt the common procedure of fitting a second order polynomial to the
upper RGB of the cluster under consideration, so that it is parametrized by 
$(V-I) = a_0 + a_1V + a_2V^2$. We fit this relation iteratively, discarding outlying
points on each loop. The results may be seen for ESO 121-SC03 and Pal. 12 in 
Fig. \ref{f:rgbfits} and Table \ref{t:metred}. Both clusters (but particularly
Pal. 12) are sparse, so their upper RGBs are not very well defined. 
This means that the resulting parametrizations are correspondingly uncertain.

Fig. \ref{f:rgbfits} and Table \ref{t:metred} also display the calculated HB levels.
In both clusters the HBs are narrow and short, so the HB levels are well defined
and straightforward to measure.
We find $V_{{\rm HB}} = 18.90 \pm 0.03$ for ESO 121-SC03, and 
$V_{{\rm HB}} = 17.05 \pm 0.03$ for Pal. 12. These values compare reasonably well
with those found in previous studies. For example, \citet{mateo} measured
$V_{{\rm HB}} = 19.0$ for ESO 121-SC03, while \citet{sarajedini:95} calculated
$V_{{\rm HB}} = 18.96$ from the same photometry. Similarly, \citet{stetson}
measured $V_{{\rm HB}} = 17.13$ for Pal. 12, while \citet{sarajedini:95} found
$V_{{\rm HB}} = 17.10$, again from the same data. \citet{gratton} obtained 
$V_{{\rm HB}} = 17.05 \pm 0.02$ for Pal. 12, and more recently \citet{rosenberg} 
measured $V_{{\rm HB}} = 17.18$ for this cluster, although they found that their 
photometric calibration in the $V$-band was on average $0.12$ mag fainter than 
that of \citet{gratton}, and $0.05$ mag fainter than that of \citet{stetson}. 
There is a tendency for our measurements to be $\sim 0.05$ mag brighter than
the literature values. It is not clear whether this is a systematic offset
or whether it simply represents random observational errors. Nonetheless, the  
consistency between our aperture photometry and that derived using {\sc dolphot} 
(see Section \ref{s:obsred}), as well as our measurements of 47 Tuc (see Section
\ref{s:ages}, below) leads us to have confidence in our photometry. For the
most part, any such small global offsets will in any case be irrelevant, since 
from here on we mostly consider differential measures on the CMD (e.g., the 
difference between $V_{{\rm HB}}$ and the main-sequence turn-off level 
as an age indicator).

Applying Sarajedini's (1994) method, we obtained metallicity and reddening 
measurements of $[$Fe$/$H$] = -0.97 \pm 0.10$ and $E(V-I) = 0.04 \pm 0.02$ 
for ESO 121-SC03, and $[$Fe$/$H$] = -1.05 \pm 0.10$ and $E(V-I) = 0.00 \pm 0.02$
for Pal. 12. These are recorded in Table \ref{t:metred}.

Our measurements for ESO 121-SC03 are in good agreement with previous
results. \citet{mateo} concluded $[$Fe$/$H$] = -0.9 \pm 0.2$ for this
cluster from their $(B,\,B-V)$ CMD, while \citet{olszewski} measured 
$[$Fe$/$H$] = -0.93 \pm 0.1$ spectroscopically and \citet{bica} derived
$[$Fe$/$H$] = -1.05 \pm 0.2$ from Washington photometry. More recently, \citet{hill}
obtained $[$Fe$/$H$] = -0.91 \pm 0.16$ from high resolution spectroscopic
measurements made with VLT/UVES. There are fewer estimates of the foreground
reddening for ESO 121-SC03. Both \citet{mateo} and \citet{bica} adopt 
$E(B-V) = 0.03$ from the \citet{burstein} maps, which corresponds to 
$E(V-I) \sim 0.04$ if we use the relation that $E(V-I) = 1.31E(B-V)$
\citep[see e.g.,][]{mackeyred}. For comparison, the \citet{schlegel} dust maps 
give a value of $E(B-V) = 0.04$, which is also consistent with our measurement.
The close agreement between our present results and those in the literature
suggests that we have not introduced any significant systematic errors into
our photometry by transforming out of the native ACS/WFC photometric system.

For Pal. 12, a number of previous metallicity estimates exist. As summarized by
\citet{rosenberg}, \citet{dacosta:90} determined $[$Fe$/$H$] = -1.06 \pm 0.12$
from $VI$ photometry, while \citet{arman:91} measured $[$Fe$/$H$] = -0.60 \pm 0.14$
spectroscopically, and \citet{dacosta:95} found $[$Fe$/$H$] = -0.64 \pm 0.09$ from a
re-analysis of the same data. \citet{brown} conducted a higher resolution spectroscopic
study of two Pal. 12 giants, obtaining $[$Fe$/$H$] = -1.0 \pm 0.1$, and 
\citet{rosenberg} themselves measured $[$Fe$/$H$] \simeq -0.93$ from $VI$
photometry. Most recently, \citet{cohen} has measured $[$Fe$/$H$] \simeq -0.8$
from high resolution Keck spectra of four Pal. 12 giants. For the reddening
towards this cluster, \citet{rosenberg} assume $E(V-I) = 0.03$ based on several
earlier measurements, while the \citet{schlegel} maps imply $E(V-I) \simeq 0.05$.
\citet[][2003 update]{harris} lists $E(B-V) = 0.02$, which corresponds to
$E(V-I) \simeq 0.03$.

Our measured metallicity and reddening for Pal. 12 do not agree as closely
with previous measurements as our results for ESO 121-SC03 do; however we feel that
this is most likely due to the poor definition of the Pal. 12 RGB on our CMD rather 
than any large systematic error in the photometry. Pal. 12 is a sparse, diffuse cluster, 
and this is compounded by the relatively small ACS/WFC field of view -- \citet{rosenberg} 
find a tidal radius of $r_t \sim 7.6\arcmin$, compared with the 
$\sim 3.3\arcmin \times 3.3\arcmin$ ACS/WFC field of view. Hence we have few stars with 
which to determine the RGB fiducial. In particular, Fig. \ref{f:rgbfits} shows that
the uncertainty in the colour of the RGB at the HB level is relatively large, which
explains the small discrepancy between our foreground reddening estimate and those 
in the literature. Similarly, the upper RGB is effectively defined by only two
stars, which introduces a corresponding uncertainty into our metallicity estimate.

Given the above results and discussion, we will adopt $[$Fe$/$H$] = -0.95 \pm 0.05$ and
$E(V-I) = 0.04 \pm 0.02$ for ESO 121-SC03 for the remainder of this paper. For Pal. 12
we will assume $[$Fe$/$H$] = -0.80 \pm 0.10$ based on the recent high quality spectra of
\citet{cohen}\footnote{We note that the mean of the six literature measurements discussed
above, plus our own for this cluster is $[$Fe$/$H$] = -0.84$, in good agreement
with the \citet{cohen} measurement.}, along with a colour excess $E(V-I) = 0.03 \pm 0.02$. 

\subsection{Distance of ESO 121-SC03}
\label{ss:distance}
It is worth briefly considering the distance of ESO 121-SC03, both from us and from
the centre of the LMC. In the previous Section we measured the HB level for this cluster
to be $V_{{\rm HB}} = 18.90 \pm 0.03$. We also estimated a colour excess
$E(V-I) = 0.04 \pm 0.02$, which corresponds to a $V$-band extinction 
$A_V = 0.095 \pm 0.047$. 
To determine the distance modulus, we must adopt some relationship between metallicity
and the intrinsic luminosity of the HB. \citet{chaboyer} concluded that
\begin{equation}
M_V({\rm HB}) = (0.23 \pm 0.04)([{\rm Fe}/{\rm H}] + 1.6) + (0.56 \pm 0.12) \,\,,
\end{equation}
while \citet{gratton:03} found that 
\begin{equation}
M_V({\rm HB}) = (0.22 \pm 0.05)([{\rm Fe}/{\rm H}] + 1.5) + (0.56 \pm 0.07) \,\,,
\end{equation}
from their detailed study of the globular clusters NGC 6397, NGC 6752 and 47 Tuc.
Taking the mean of the two values of $M_V({\rm HB})$ so calculated for ESO 121-SC03
(with $[$Fe$/$H$] = -0.95 \pm 0.05$), we find $M_V({\rm HB}) = 0.695 \pm 0.010$ for 
this cluster (where the quoted error is due only to our random error in $[$Fe$/$H$]$).
In combination with our measured HB level and $V$-band extinction this implies a
distance modulus $\mu = 18.11 \pm 0.09$. This is significantly shorter than the
standard LMC distance modulus $\mu_{{\rm LMC}} = 18.50 \pm 0.09$ 
\citep[see e.g.,][ and references therein]{gratton:03}. In terms of linear distance,
ESO 121-SC03 is $\sim 20$ per cent closer to us than is the LMC.

This result suggests that ESO 121-SC03 is located further away from the centre of the 
LMC than its projected angular separation would imply. Adopting the optical centre
of the LMC from \citet{bica:lmc}, at 
$\alpha = 05^{{\rm h}}\,20^{{\rm m}}\,56^{{\rm s}}$, 
$\delta = -69\degr\,28\arcmin\,41\arcsec$,
ESO 121-SC03 lies at a projected angular separation of $9.9\degr$. Assuming the two
distance moduli quoted above leads to the conclusion that the linear distance between
the centre of the LMC and ESO 121-SC03 is close to $11.5$ kpc. This renders 
ESO 121-SC03 one of the most remote known LMC star clusters.

\subsection{Blue stragglers in ESO 121-SC03}
\label{ss:bss}
As noted above, our CMD for ESO 121-SC03 shows the population of blue straggler
candidates to be rather large, including a significant number of 
very blue objects. Since we know of no prior discussion of blue stragglers in 
ESO 121-SC03, we examined this population in more detail.

\begin{figure}
\includegraphics[width=0.5\textwidth]{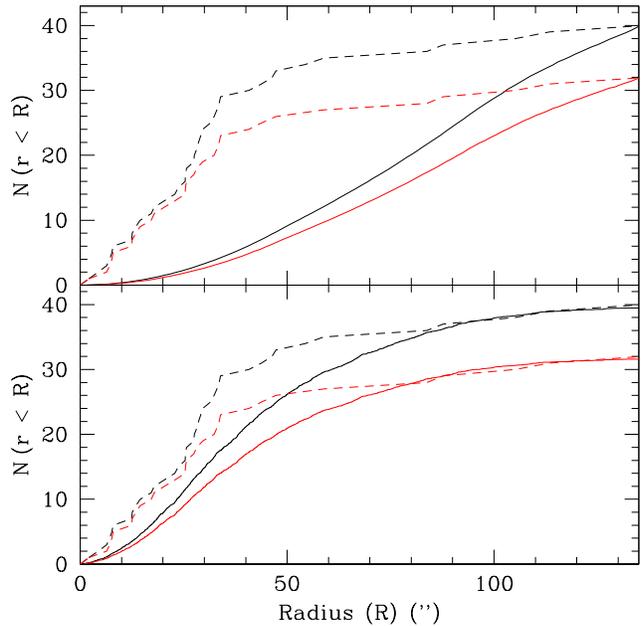}
\caption{Cumulative distributions of blue straggler candidates with projected radius.
In both panels, the dashed lines represent the observed distributions for our cleaned sample 
of $32$ (lower line) and our original sample of $40$ (upper line). In the upper panel, for 
comparison we have also plotted the expected distributions for a uniformly distributed 
background with surface density normalized as described in the text (solid lines). 
Clearly our blue straggler candidates are associated with the cluster rather than a uniform 
field star population. The abrupt flattening in the blue straggler distributions beyond 
$\sim 50\arcsec$ is at least partly due to the restricted ACS/WFC field of view. This flattening
is also visible in the solid lines, which deviate from a pure $R^2$ relation at large radii.
In the lower panel the solid lines represent the radial distribution of upper main
sequence stars (see text). The blue straggler population is clearly more centrally concentrated
than these main sequence objects.}
\label{f:cdbss}
\end{figure}

In Fig. \ref{f:cmdvimag} we have selected blue straggler candidates by defining
a region on the CMD which encompasses them. The number of stars falling within this
region is $40$. We examined each star by eye on our F555W and F814W ACS/WFC images
of ESO 121-SC03, and eliminated all stars which showed some hint of being affected by
blends or cosmic ray strikes. This reduced the sample to $32$ stars. These are marked
in bold on Fig. \ref{f:cmdvimag}. Clearly, the large and extended blue straggler 
population remains, so we are confident that our photometry for these stars is good.

Since we have not made any attempt to remove field stars from our CMD, we would like
to rule out the possibility of significant contamination by such objects. We therefore
calculated the projected radius of each candidate blue straggler from the cluster
centre, and constructed a cumulative distribution of the number of blue stragglers with
projected radius. This may be seen in the upper panel of Fig. \ref{f:cdbss}. In this plot, 
the two dashed lines represent the cumulative distributions for our original sample of $40$
stars and our cleaned sample of $32$. The two solid lines show the expected cumulative
distribution with radius of a uniform background stellar population with a surface
density normalized to match the observed number of blue straggler candidates within the 
maximum radius (either $40$ or $32$). Note that in constructing these two comparison lines,
we have accounted for the ACS/WFC field-of-view -- hence the deviation from a pure $R^2$
function visible at larger radii. From this plot, it is clear that the blue straggler 
candidates are far more centrally concentrated than a uniformly distributed background, which 
strongly suggests they are cluster members rather than field stars.

In the lower panel of Fig. \ref{f:cdbss} we compare the radial distribution of blue stragglers
with that of main sequence stars in ESO 121-SC03. We selected our main sequence sample
by taking all stars below the main-sequence turn-off (at $V = 22.16$ -- see Section 
\ref{ss:vertical}), above the approximate faint detection limit (at $V \sim 26$), and
within $3\sigma$ of the approximate main-sequence ridge-line. This sample does not represent
all main-sequence stars, since our photometry is not deep enough to reach very low-mass
objects; but is representative of the most massive present-day main sequence members.
The cumulative distributions of these stars, normalized to our two blue straggler sample 
sizes, are plotted as the solid lines in the lower panel of Fig. \ref{f:cdbss}. Clearly the 
blue straggler candidates are significantly more centrally concentrated than the main sequence
stars. As demonstrated below, this observation is consistent with dynamical mass segregation
having occurred in ESO 121-SC03.

\begin{table*}
\begin{minipage}{164mm}
\caption{ACS/WFC observations of 47 Tuc (Program 9018).}
\begin{tabular}{@{}lcccccccc}
\hline \hline
Cluster & \hspace{10mm} & RA & Dec. & \hspace{10mm} & Filter & Image & Exposure & Date \\
 & & (J2000.0) & (J2000.0) & & & Name & Time (s) & \\
\hline
47 Tuc (pair 1) & & $00^{{\rm h}}\,\,22^{{\rm m}}\,\,37.20^{{\rm s}}$ & $-72\degr\,\,04\arcmin\,\,14.0\arcsec$ & & F555W & j8c0b1tzq & $30$ & $09/05/2002$ \\ 
 & & & & & F814W & j8c0d1caq & $30$ & $06/05/2002$ \\
47 Tuc (pair 2) & & $00^{{\rm h}}\,\,22^{{\rm m}}\,\,37.20^{{\rm s}}$ & $-72\degr\,\,04\arcmin\,\,14.0\arcsec$ & & F555W & j8c0b1u1q & $30$ & $09/05/2002$ \\ 
 & & & & & F814W & j8c0d1ccq & $30$ & $06/05/2002$ \\ 
47 Tuc (pair 3) & & $00^{{\rm h}}\,\,22^{{\rm m}}\,\,37.20^{{\rm s}}$ & $-72\degr\,\,04\arcmin\,\,14.0\arcsec$ & & F555W & j8c0b1skq & $30$ & $09/05/2002$ \\ 
 & & & & & F814W & j8c032xuq & $5$  & $19/04/2002$ \\ 
\hline
\label{t:47tuc}
\end{tabular}
\end{minipage}
\end{table*}

\citet{piotto} examined blue straggler relative frequency $F_{{\rm BSS}}$ as a 
function of integrated cluster luminosity $M_V$ and of central luminosity density 
$\rho_0$ for a large sample of Galactic globular clusters observed with {\it HST}. 
They found a significant anti-correlation between $F_{{\rm BSS}}$ and $M_V$, and 
also between $F_{{\rm BSS}}$ and $\rho_0$ for $\log \rho_0 < 3.2$ (see their Figure 1). 
It is instructive to consider where ESO 121-SC03 lies with regard to these
anti-correlations. $F_{{\rm BSS}}$ is defined in \citet{piotto} as the number of
blue stragglers normalized to the number of horizontal branch stars. For ESO 121-SC03,
we observe $N_{{\rm HB}} = 19$ from Fig. \ref{f:cmdvimag}. If we then assume
a population $N_{{\rm BSS}} = 36 \pm 4$, we find $F_{{\rm BSS}} = 1.9 \pm 0.2$,
i.e., $\log F_{{\rm BSS}} \approx 0.3$. 

Adopting the best fitting structural parameters of \citet{mateo}, who found
a central surface brightness $\sigma_{V,0} = 22.3$, a core radius $r_c = 34\arcsec$
and tidal radius $r_t = 136\arcsec$ for ESO 121-SC03, we estimate $\log \rho_0 \approx 0.7$ 
for this cluster according to the prescription of \citet{djorg}. \citet{mateo} also
obtained the apparent integrated magnitude of ESO 121-SC03, $V_{{\rm ESO}} = 13.4$,
which at our distance modulus and assumed reddening corresponds to $M_V \approx -4.8$.
From examination of Figure 1 in \citet{piotto}, we see that ESO 121-SC03 falls at 
much lower $\log \rho_0$ and $M_V$ than do any of the Galactic globular clusters
in their sample. It also has a higher $F_{{\rm BSS}}$ than all but two of the clusters
in their sample. Hence our observations for ESO 121-SC03 are consistent with, and 
significantly extend the anti-correlations observed by these authors.

Since stellar collisions are clearly not frequent in as low a density cluster as 
ESO 121-SC03, the observed population of blue stragglers is presumably linked to the 
evolution of primordial binary stars in this cluster. 
\citet{mateo} measure the half-mass radius of ESO 121-SC03 to be $r_h = 38\arcsec$.
Adopting a mean stellar mass of $0.33 M_\odot$ and a global mass-to-light ratio
$M/L \sim 2$, we calculate a median relaxation time for ESO 121-SC03 of $t_{rh} \sim 1.6$ 
Gyr, using the prescription of Eq. 8-72 in \citet{binney}. Since dynamical mass segregation
occurs on approximately the two-body relaxation time-scale, our observation that the blue 
stragglers are more centrally concentrated than the most massive main sequence stars in 
ESO 121-SC03 is consistent with dynamical mass segregation having occurred in this cluster, 
and with the likelihood that the observed population of blue stragglers is linked to
the evolution of primordial binary stars. 

\section{Relative Age Measurements}
\label{s:ages}
As described in Section \ref{s:intro}, ESO 121-SC03 has been shown by several studies
to be a few Gyr younger than the oldest clusters in the LMC. The availability of
our high quality ACS photometry offers the opportunity to place the tightest constraints
yet on the age of this cluster. The most precise way to do this is to obtain an
age estimate relative to other globular clusters for which well determined
measurements already exist. One such cluster is Pal. 12, which is why we have included
it in the present study. As discussed previously, this cluster has been demonstrated
by a number of authors to be considerably younger than the oldest Galactic globular
clusters, just as ESO 121-SC03 is younger than the oldest LMC objects. For example, 
\citet{rosenberg} found Pal. 12 to have an age of only $68 \pm 10$ per cent that of
47 Tuc and M5. 

Ideally, we would like to also include an older globular cluster than ESO 121-SC03
in the relative age measurements. Only one such cluster, of similar metallicity to 
ESO 121-SC03, has been observed with ACS/WFC in the F555W and F814W filters. This is 
47 Tuc (NGC 104), with $[$Fe$/$H$] = -0.76$ \citep{harris}. In order to include this 
object in our present study, we obtained suitable imaging from the HST archive 
(taken as part of the ACS calibration program
9018) and reduced it using exactly the same photometric pipeline (described in 
Section \ref{s:obsred}) that we used for ESO 121-SC03 and Pal. 12. The three pairs 
of specific images we used, and their exposure durations, are listed in Table 
\ref{t:47tuc}. Note that these observations are centred on a field located 
$\sim 6.8\arcmin$ from the nominal cluster centre (which is at 
$00^{{\rm h}}\,24^{{\rm m}}\,05.2^{{\rm s}}$, $-72\degr\,04\arcmin\,51.0\arcsec$).
This offset helped ameliorate any crowding issues associated with observing near
the centre of this object.

\begin{figure}
\includegraphics[width=0.5\textwidth]{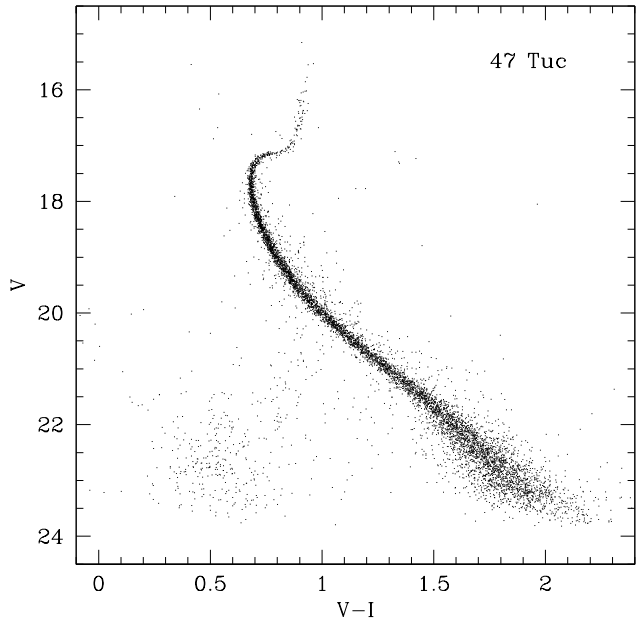}
\caption{Cleaned $VI$ colour-magnitude diagram for 47 Tuc. As described in the text, these
measurements are derived from observations of a field offset from the cluster centre
by $\sim 6.8\arcmin$. The non-cluster population evident around $(V,\,V-I) = (23,\,0.5)$
are background SMC field stars.}
\label{f:47tuc}
\end{figure}

At the end of the photometric reduction we were left with three CMDs from the three 
pairs of observations. We matched stellar coordinates between these lists, keeping 
only objects appearing in all three. The measured magnitudes were combined in an error 
weighted mean to obtain the final photometric measurements. The final, transformed 
CMD may be seen in Fig. \ref{f:47tuc}, consisting of a long, narrow main sequence and 
very well defined turn-off region. Unfortunately, because of the image exposure 
durations, all stars brighter than $V \sim 15.5$ are very saturated and could not 
be measured accurately. No shorter duration images are available in the archive, so
we could not obtain a CMD including the HB. Nevertheless, this incompleteness did
not compromise our age measurements. As discussed below, 47 Tuc is a very 
well studied cluster and it is possible to obtain good estimates of the HB level from
the literature.

\subsection{Theoretical calibration}
\label{ss:theory}
For our relative age measurements, we used two techniques -- the so-called
vertical and horizontal methods. Both employ differential indicators on the CMD.
The vertical method relies on the fact that the difference between the level of
the HB ($V_{{\rm HB}}$) and the level of the MSTO ($V_{{\rm TO}}$) is age-dependent,
with older clusters having a larger value of this parameter. Similarly, the horizontal
method makes use of the difference in colour between the MSTO and some point on the
RGB below the HB. This relies on the fact that older clusters generally have shorter
sub-giant branches (SGBs) and hence bluer RGBs below the HB. 

Both the vertical and horizontal age dating techniques have some metallicity dependence 
which must be accounted for, which is one reason why we have selected reference 
clusters of similar metallicity to ESO 121-SC03. There is also some dependence on 
the abundance of $\alpha$-process elements (e.g., O, Mg, Si, Ca, and Ti), which we 
would like to account for. It is well known that many Galactic globular clusters are 
enhanced in $\alpha$-process elements relative to the solar value, with 
$[\alpha/$Fe$] \sim +0.3$. A similar over-abundance has been observed for some old LMC 
clusters \citep[see e.g.,][]{johnson,puzia}. However, we know of only one relevant measurement 
for ESO 121-SC03 -- that of \citet{hill}, who observed $[$O$/$Fe$] = +0.15$ in two
red giants from high resolution spectra. These authors also measured $[$Al$/$Fe$]$ in their
two giants, and from the low observed abundance ($\sim -0.4$) suggest that deep mixing
has not occurred and therefore that the observed oxygen abundance is genuinely lower
than that in comparable Galactic globular cluster giants, which typically have 
$[$O$/$Fe$] \sim +0.3-0.4$. 

Apart from this measurement, the $\alpha$-abundance of ESO 121-SC03 remains essentially 
unknown. Fortunately, our reference clusters span the range of $[\alpha/$Fe$]$
within which ESO 121-SC03 is likely to lie, as indicated by the observed value of $[$O$/$Fe$]$. 
47 Tuc possesses the enhancement typical of many Galactic globulars -- for example, from 
\citet{carney} we infer $[\alpha/$Fe$] \simeq 0.18$ for this cluster, while \citet{gratton:03} 
list $[\alpha/$Fe$] = 0.30 \pm 0.02$. On the other hand, Pal. 12 shows no such
enhancement -- as demonstrated by the spectroscopic studies of \citet{brown}
and \citet{cohen}, the latter of who found the mean of $[$Si$/$Fe$]$, $[$Ca$/$Fe$]$,
and $[$Ti$/$Fe$]$ to be $-0.07 \pm 0.05$. Indeed, this pattern is one of the strongest 
clues linking Pal. 12 with the Sagittarius dwarf galaxy \citep{cohen}. 

In order to establish a theoretical calibration against which to judge the relative
ages of our three clusters, we used the recently published Victoria-Regina stellar 
models. For full details of these, see \citet{vandenberg} and the references therein. 
The set of models most relevant to us consists of $60$ grids of
stellar evolutionary tracks -- at each of $20$ $[$Fe$/$H$]$ values between
$-2.31$ and the solar metallicity, for three $[\alpha/$Fe$]$ values ($0.0$, $+0.3$, 
and $+0.6$) per given iron abundance. For each grid, isochrones may be interpolated 
in $UBVRI$ for ages between $1$ and $18$ Gyr. Also provided are zero-age HB models
which may be transformed to the observational plane at each $[$Fe$/$H$]$ and
$[\alpha/$Fe$]$ point.

For the present study we selected the grids constituting the four possible combinations
of $[$Fe$/$H$] = -0.83$ and $-1.01$, and $[\alpha/$Fe$] = 0.0$ and $+0.3$. These
cover the range appropriate for the clusters in our present study. We interpolated
$V,\,V-I$ isochrones from each grid for ages between $6$ and $17$ Gyr, at $0.5$ Gyr
intervals. On each isochrone we measured values of our vertical method and horizontal
method indicators. For the vertical method we chose to use the difference in $V$
magnitude between the level of the MSTO and the level of the HB, labelled 
$\Delta V_{{\rm TO}}^{{\rm HB}}$, while for the horizontal method we used 
$\delta (V-I)_{2.2}$ -- the $V-I$ colour difference between the MSTO and the point 
on the RGB $2.2$ mag brighter (in $V$) than the level of the MSTO. The exact point
chosen on the RGB does not matter, as long as it is not too close to the SGB or
the level of the HB \citep[see e.g.,][]{rosenberg}. For the present work we chose
the $2.2$ mag level so as to be near to the brightest RGB stars in the CMD for 47 Tuc, 
without incorporating saturated objects.

\begin{figure*}
\begin{minipage}{175mm}
\begin{center}
\includegraphics[width=87mm]{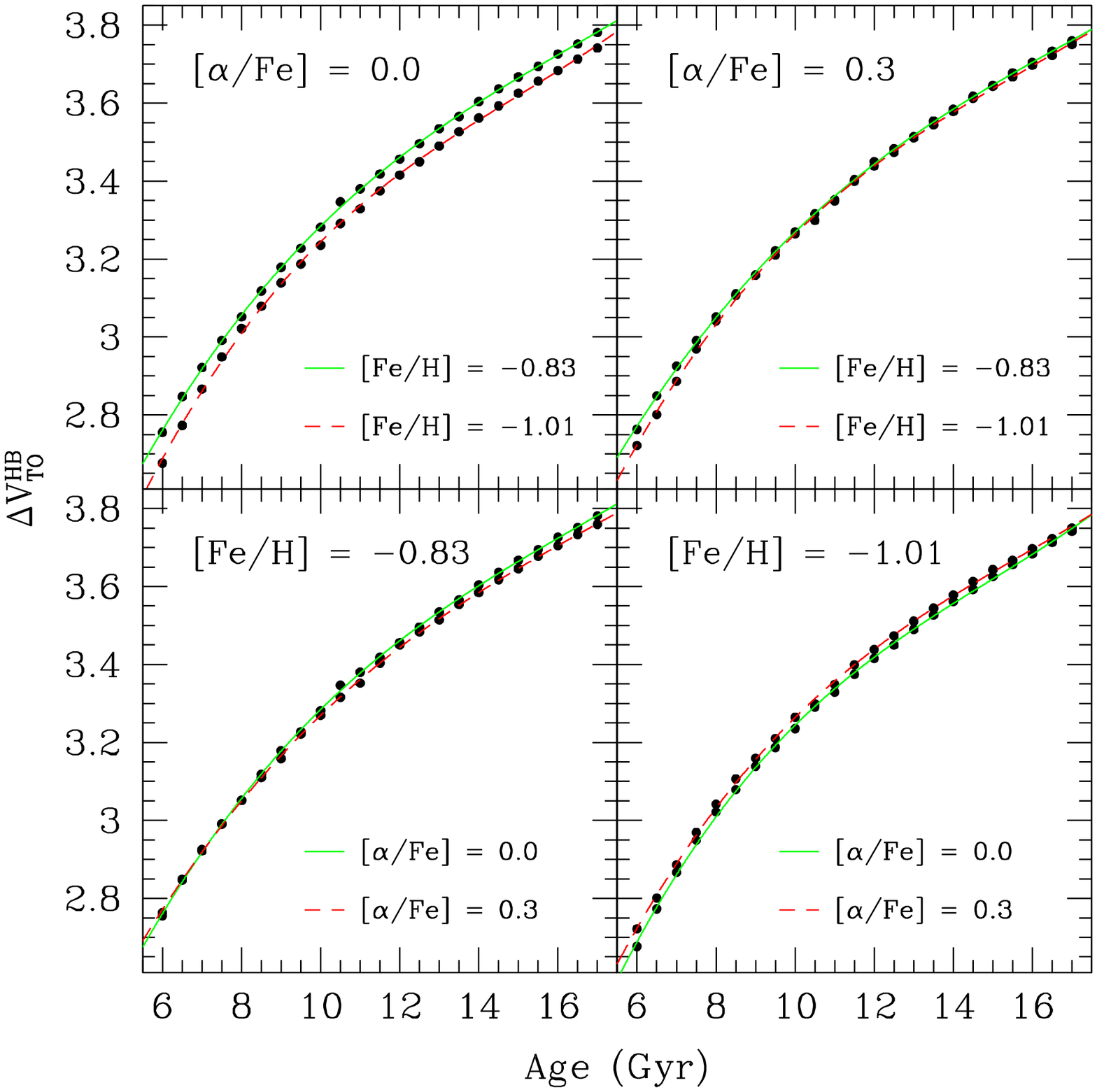}
\hspace{-2mm}
\includegraphics[width=87mm]{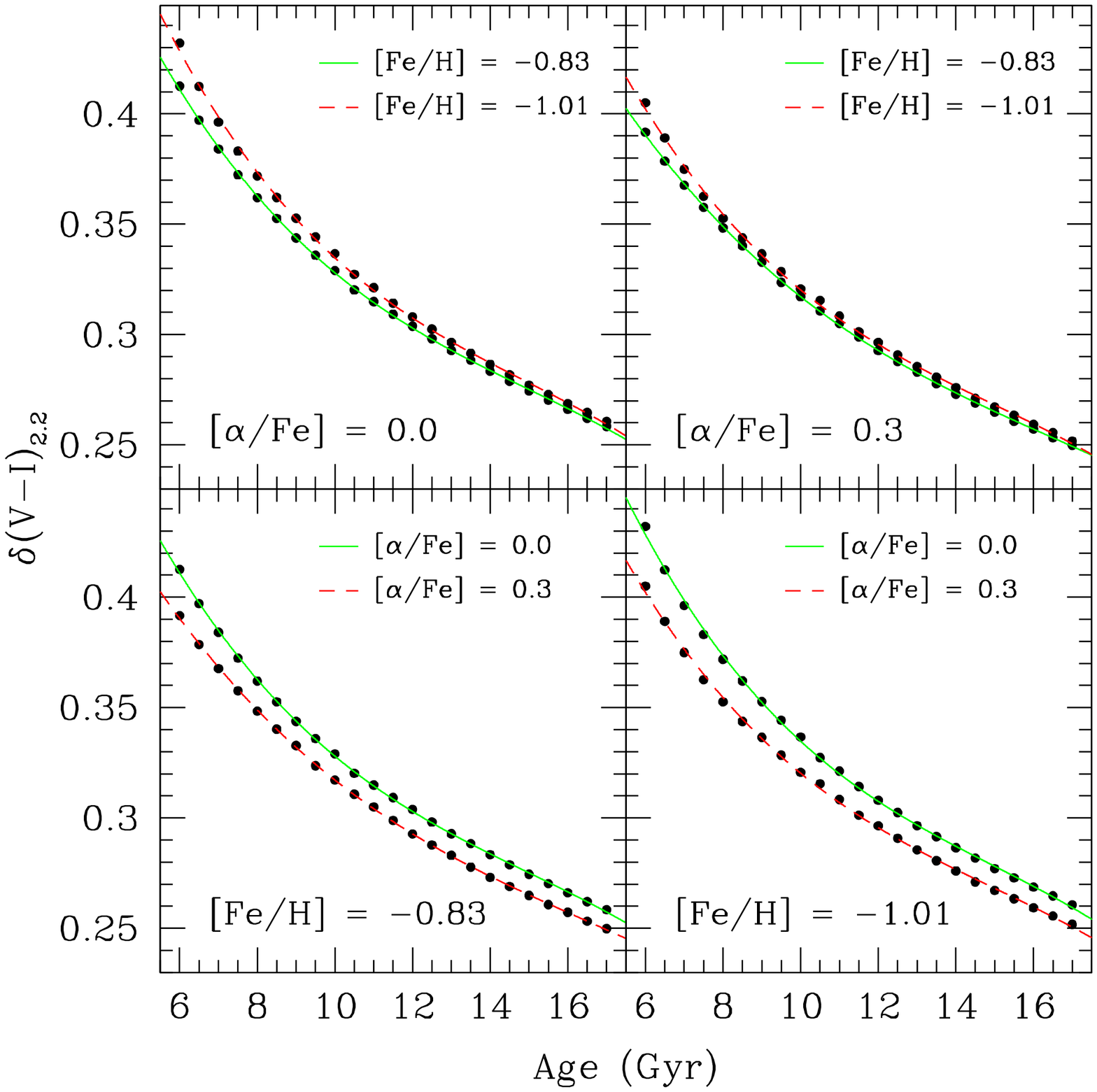}
\caption{Theoretical age calibrations for our vertical method (left) and horizontal
method (right) indicators, showing the variation with metallicity and $\alpha$-enhancement 
(solid points). Selected metallicities are $[$Fe$/$H$] = -0.83$ and $-1.01$, each 
with $[\alpha/$Fe$] = 0.0$ and $+0.3$. The green (solid) and red (dashed) lines represent 
the third-order polynomial fits, as described in the text.}
\label{f:calib}
\end{center}
\end{minipage}
\end{figure*}

\begin{table*}
\begin{minipage}{122mm}
\caption{Vertical and horizontal method age calibration coefficients.}
\begin{tabular}{@{}lcccccccc}
\hline \hline
Method & \hspace{3mm} & $[$Fe$/$H$]$ & $[\alpha/$Fe$]$ & \hspace{3mm} & $b_0$ & $b_1$ & $b_2$ & $b_3$ \\
\hline
Vertical & & $-0.832$ & $+0.00$ & & $-63.2574$ & $67.7981$ & $-24.0045$ & $3.09148$ \\
Vertical & & $-1.009$ & $+0.00$ & & $-5.40693$ & $14.9235$ & $-7.98562$ & $1.49592$ \\
Vertical & & $-0.832$ & $+0.30$ & & $-70.1512$ & $73.6266$ & $-25.7156$ & $3.27090$ \\
Vertical & & $-1.009$ & $+0.30$ & & $-56.1248$ & $63.0419$ & $-23.1402$ & $3.07405$ \\
 & & & & & & & &  \\
Horizontal & & $-0.832$ & $+0.00$ & & $103.963$ & $-600.912$ & $1261.43$ & $-922.159$ \\
Horizontal & & $-1.009$ & $+0.00$ & & $100.151$ & $-569.686$ & $1185.40$ & $-859.686$ \\
Horizontal & & $-0.832$ & $+0.30$ & & $111.519$ & $-700.206$ & $1619.40$ & $-1327.04$ \\
Horizontal & & $-1.009$ & $+0.30$ & & $98.2591$ & $-565.779$ & $1178.87$ & $-850.567$ \\
\hline
\label{t:agecalib}
\end{tabular}
\end{minipage}
\end{table*}

To find the MSTO on each isochrone, we fit a second-order polynomial to the $\sim 10$
model points near the TO, and found its turning point. With this computed, it was
straightforward to measure the colour of the RGB at $V_{2.2} = V_{{\rm TO}} - 2.2$
and determine $V_{{\rm HB}}$ from the appropriate ZAHB model. From these measurements
we calibrated our relative age indicators as a function of $[$Fe$/$H$]$ and
$[\alpha/$Fe$]$. The results may be seen in Fig. \ref{f:calib}. In order to allow
accurate relative age determinations for our three clusters, we fit third-order polynomials 
to the measured theoretical calibrations. The best-fitting polynomials are also drawn 
in Fig. \ref{f:calib}, and their coefficients are listed in Table \ref{t:agecalib}.
These are defined by parametrizing the relationship between age, $\tau$, and 
observed indicator, $\delta$, by $\tau = b_0 + b_1 \delta + b_2 \delta^2 + b_3 \delta^3 \,$.

From Fig. \ref{f:calib}, it can be seen that the vertical indicator 
$\Delta V_{{\rm TO}}^{{\rm HB}}$ is relatively insensitive to changes in $[$Fe$/$H$]$ 
and $[\alpha/$Fe$]$ over the ranges spanned by our clusters. The greatest change in 
derived age for measured $\Delta V_{{\rm TO}}^{{\rm HB}}$ occurs with $[$Fe$/$H$]$ at 
$[\alpha/$Fe$] = 0.0$. In this case, $\Delta \tau \simeq 0.5 - 1$ Gyr moving from 
$[$Fe$/$H$] = -0.83$ to $[$Fe$/$H$] = -1.01$. 
The horizontal indicator exhibits far greater variance, particularly with $[\alpha/$Fe$]$ 
at given $[$Fe$/$H$]$. For example, at $[$Fe$/$H$] = -1.01$, moving from 
$[\alpha/$Fe$] = 0.0$ to $[\alpha/$Fe$] = +0.3$ leads to a reduction in the age of 
$\sim 10 - 15$ per cent.

Given these observations and the dearth of high resolution spectral measurements of stars
in ESO 121-SC03, we decided it prudent to consider both the case where this cluster has 
$[\alpha/$Fe$] \simeq 0.0$ and the case where it has $[\alpha/$Fe$] \simeq +0.3$. 

\begin{table*}
\begin{minipage}{169mm}
\caption{Observed vertical and horizontal relative age indicators for ESO 121-SC03, Pal. 
12, and 47 Tuc.}
\begin{tabular}{@{}lccccccccc}
\hline \hline
Cluster & \hspace{4mm} & $V_{{\rm TO}}$ & $(V-I)_{{\rm TO}}$ & \hspace{4mm} & $V_{{\rm HB}}$ & $\Delta V_{{\rm TO}}^{{\rm HB}}$ & \hspace{4mm} & $(V-I)_{2.2}$ & $\delta (V-I)_{2.2}$ \\
\hline
ESO 121-SC03 & & $22.16 \pm 0.05$ & $0.606 \pm 0.005$ & & $18.90 \pm 0.03$ & $3.26 \pm 0.06$ & & $0.941 \pm 0.005$ & $0.336 \pm 0.007$ \\
Palomar 12 & & $20.37 \pm 0.05$ & $0.613 \pm 0.005$ & & $17.05 \pm 0.03$ & $3.32 \pm 0.06$ & & $0.929 \pm 0.005$ & $0.317 \pm 0.007$ \\
47 Tuc & & $17.66 \pm 0.05$ & $0.681 \pm 0.005$ & & $14.10 \pm 0.03$ & $3.56 \pm 0.06$ & & $0.960 \pm 0.005$ & $0.279 \pm 0.007$ \\
\hline
\label{t:obsage}
\end{tabular}
\end{minipage}
\end{table*}

\subsection{Results: vertical method}
\label{ss:vertical}
On each of the three cluster CMDs, we determined the colour and level of the MSTO by 
fitting a second-order polynomial to the data in this region, and finding the turning
point of the fit. The results are listed in Table \ref{t:obsage}. We estimate our
measured values of $V_{{\rm TO}}$ are accurate to $\sim 0.05$ mag, while those for
$(V-I)_{{\rm TO}}$ are accurate to $\sim 0.005$ mag.

\begin{table*}
\begin{minipage}{161mm}
\caption{Relative age results -- vertical method.}
\begin{tabular}{@{}lcccclccccc}
\hline \hline
Cluster 1 & $[$Fe$/$H$]$ & $[\alpha/$Fe$]$ & $\Delta V_{{\rm TO}}^{{\rm HB}}$ & \hspace{4mm} & Cluster 2 & $[$Fe$/$H$]$ & $[\alpha/$Fe$]$ & $\Delta V_{{\rm TO}}^{{\rm HB}}$ & \hspace{4mm} & $\tau_1 / \tau_2$ \\
\hline
ESO 121-SC03 & $-1.009$ & $+0.00$ & $3.26 \pm 0.06$ & & 47 Tuc & $-0.832$ & $+0.30$ & $3.56 \pm 0.06$ & & $0.75 \pm 0.09$ \\
ESO 121-SC03 & $-1.009$ & $+0.30$ & $3.26 \pm 0.06$ & & 47 Tuc & $-0.832$ & $+0.30$ & $3.56 \pm 0.06$ & & $0.73 \pm 0.09$ \\
 & & & & & & & & & & \\
ESO 121-SC03 & $-1.009$ & $+0.00$ & $3.26 \pm 0.06$ & & Pal. 12 & $-0.832$ & $+0.00$ & $3.32 \pm 0.06$ & & $0.99 \pm 0.11$ \\
ESO 121-SC03 & $-1.009$ & $+0.30$ & $3.26 \pm 0.06$ & & Pal. 12 & $-0.832$ & $+0.00$ & $3.32 \pm 0.06$ & & $0.97 \pm 0.11$ \\
 & & & & & & & & & & \\
Pal. 12 & $-0.832$ & $+0.00$ & $3.32 \pm 0.06$ & & 47 Tuc & $-0.832$ & $+0.30$ & $3.56 \pm 0.06$ & & $0.76 \pm 0.09$ \\
\hline
\label{t:agesvert}
\end{tabular}
\end{minipage}
\end{table*}

To calculate the vertical age indicator $\Delta V_{{\rm TO}}^{{\rm HB}}$, for ESO 121-SC03
and Pal. 12 we adopted $V_{{\rm HB}}$ as measured in Section \ref{ss:metallicity}. 
We are confident that these values are close indicators of the ZAHB level in these
two clusters -- for example, \citet{recio} recommend adopting the level $3\sigma$ above
the lower envelope of the HB for clusters more metal-rich than $[$Fe$/$H$] = -1$, where
$\sigma$ is the typical error in $V$ at the HB magnitude.

For 47 Tuc we do not have photometry of the HB region, so we searched the literature
for recent accurate measurements of the HB level in this cluster. In their recent
relative age study of $55$ Galactic globular clusters, \citet{deangeli} measured 
$V_{{\rm HB}}$ for 47 Tuc from both ground-based and HST/WFPC2 imaging, using the
prescription of \citet{recio}. They determined
$V_{{\rm HB}} = 14.10 \pm 0.03$, which is the value we adopt for the present work.
We are confident their photometry is on a very similar scale to ours for 47 Tuc, as they 
also measured $V_{{\rm TO}} = 17.65 \pm 0.08$, which is almost identical to our value.
We note that \citet{harris} lists $V_{{\rm HB}} = 14.06 \pm 0.10$ for 47 Tuc, consistent
with the measurement adopted here.

Next, we applied our theoretical age calibration to the measured values of 
$\Delta V_{{\rm TO}}^{{\rm HB}}$ to determine relative ages for the three clusters.
The results may be seen in Table \ref{t:agesvert}. In these calculations, we adopted
$[$Fe$/$H$] \simeq -0.8$ for Pal. 12, as discussed in Section \ref{ss:metallicity}. 
That is, we used the $[$Fe$/$H$] = -0.83$ and $[\alpha/$Fe$] = 0.0$ polynomial
fit (see Table \ref{t:agecalib}) for this cluster. For 47 Tuc we used the 
$[$Fe$/$H$] = -0.83$ and $[\alpha/$Fe$] = +0.3$ polynomial fit, while for ESO 121-SC03
we adopted the polynomials with $[$Fe$/$H$] = -1.01$ and $[\alpha/$Fe$] = 0.0$ or
$+0.3$. 

As suspected from close examination of Fig. \ref{f:calib}, we found that varying
the $\alpha$-element abundance for ESO 121-SC03 does not result in large variations
of the relative ages derived for this cluster via the vertical method. ESO 121-SC03
is $74 \pm 9$ per cent as old as 47 Tuc, and almost identical in age
to Pal. 12. In order to compare our results with previous work, we also derived the
age of Pal. 12 relative to 47 Tuc. We found that Pal. 12 is $76 \pm 9$ per cent
the age of 47 Tuc, which is consistent with previous measurements. For example, 
\citet{rosenberg} found Pal. 12 to have an age $68 \pm 10$ per cent that of 47 Tuc,
using a horizontal indicator and several different sets of theoretical calibrations.

In order to place our relative age measurements on an absolute scale, we consider previous
age determinations for 47 Tuc. One of the most recent studies, that of \citet{gratton:03},
derived an age $\sim 2.6$ Gyr younger than NGC 6397 and NGC 6752, which apparently represent
the oldest population in the Galactic globular cluster system, with ages 
$13.4 \pm 0.8 \pm 0.6$ Gyr (where the first error bar accounts for random effects, and
the second for systematic errors). According to this estimate, 47 Tuc has an age
$\sim 81$ per cent that of the oldest Galactic globulars. This is reasonably consistent
with the relative age studies of \citet{rosenberg:99}, who found an age $\sim 90$ per cent
those of NGC 6937 and NGC 6752, and \citet{salaris}, who found an age $\sim 87$ per cent
those of NGC 6937 and NGC 6752. In contrast, the recent relative age study of
\citet{deangeli} did not find 47 Tuc to be any younger than the oldest Galactic globulars
(including NGC 6397 and NGC 6752), using both HST/WFPC2 and ground-based photometry.

If the oldest Galactic globular clusters are indeed $\sim 13.4$ Gyr old, and 47 Tuc
is $\sim 15$ per cent younger than these objects, then we find ESO 121-SC03 must
have an absolute age of $8.4 \pm 1.0$ Gyr, according to our vertical method measurements. 
Similarly, Pal. 12 must be $8.6 \pm 1.0$ Gyr old. Conversely, if 47 Tuc is just as old as 
the oldest Galactic globulars, then ESO 121-SC03 is $9.9 \pm 1.2$ Gyr old, and Pal. 12
is $10.2 \pm 1.2$ Gyr old. These results for ESO 121-SC03 are entirely consistent with
the age estimates of \citet{mateo}, who found this cluster to be $10 \pm 2$ Gyr old
if the LMC distance modulus is $18.2$, or $8 \pm 2$ Gyr old if the LMC distance
modulus is $18.7$. Similarly, \citet{bica} found an age of $8.5$ Gyr from Washington
photometry.

\subsection{Results: horizontal method}
\label{ss:horizontal}
On each of the three cluster CMDs we located the point on the RGB $2.2$ mag brighter
than our measured MSTO level, and estimated the colour by eye. Given the narrowness of
the RGB sequences below the HB level, it was straightforward to obtain these measurements
with an accuracy better than $0.01$ mag. In fact, via some experimentation we estimate
errors of $\pm 0.005$ mag in this colour. We then calculated $\delta (V-I)_{2.2}$.
These results are listed in Table \ref{t:obsage}.

\begin{table*}
\begin{minipage}{167mm}
\caption{Relative age results -- horizontal method.}
\begin{tabular}{@{}lcccclccccc}
\hline \hline
Cluster 1 & $[$Fe$/$H$]$ & $[\alpha/$Fe$]$ & $\delta (V-I)_{2.2}$ & \hspace{4mm} & Cluster 2 & $[$Fe$/$H$]$ & $[\alpha/$Fe$]$ & $\delta (V-I)_{2.2}$ & \hspace{4mm} & $\tau_1 / \tau_2$ \\
\hline
ESO 121-SC03 & $-1.009$ & $+0.00$ & $0.336 \pm 0.007$ & & 47 Tuc & $-0.832$ & $+0.30$ & $0.279 \pm 0.007$ & & $0.75 \pm 0.07$ \\
ESO 121-SC03 & $-1.009$ & $+0.30$ & $0.336 \pm 0.007$ & & 47 Tuc & $-0.832$ & $+0.30$ & $0.279 \pm 0.007$ & & $0.68 \pm 0.07$ \\
 & & & & & & & & & & \\
ESO 121-SC03 & $-1.009$ & $+0.00$ & $0.336 \pm 0.007$ & & Pal. 12 & $-0.832$ & $+0.00$ & $0.317 \pm 0.007$ & & $0.92 \pm 0.08$ \\
ESO 121-SC03 & $-1.009$ & $+0.30$ & $0.336 \pm 0.007$ & & Pal. 12 & $-0.832$ & $+0.00$ & $0.317 \pm 0.007$ & & $0.83 \pm 0.08$ \\
 & & & & & & & & & & \\
Pal. 12 & $-0.832$ & $+0.00$ & $0.317 \pm 0.007$ & & 47 Tuc & $-0.832$ & $+0.30$ & $0.279 \pm 0.007$ & & $0.81 \pm 0.08$ \\
\hline
\label{t:ageshoriz}
\end{tabular}
\end{minipage}
\end{table*}

Next, we applied our theoretical age calibration in order to obtain relative age 
estimates from the horizontal method. The results may be seen in Table \ref{t:ageshoriz}.
As previously, we used the $[$Fe$/$H$] = -0.83$ and 
$[\alpha/$Fe$] = 0.0$ polynomial fit for Pal. 12; the $[$Fe$/$H$] = -0.83$ and 
$[\alpha/$Fe$] = +0.3$ polynomial fit for 47 Tuc; and the polynomials with 
$[$Fe$/$H$] = -1.01$ and $[\alpha/$Fe$] = 0.0$ or $+0.3$ for ESO 121-SC03. 

From Table \ref{t:ageshoriz}, it can be seen that the horizontal indicator is apparently
more sensitive to $\alpha$-element abundance than is the vertical indicator, although
we note that the supposed variation with $\alpha$-abundance is of the same order as
the quoted errors. Overall, using the horizontal indicator we found ESO 121-SC03 to
have an age $72 \pm 7$ per cent that of 47 Tuc, in good agreement with our result from 
the vertical indicator. Similarly, ESO 121-SC03 has an age $88 \pm 8$ per cent
that of Pal. 12. For comparison, we find Pal. 12 to be $81 \pm 8$ per cent of the 
age of 47 Tuc, which is again in agreement with the relative age we derived 
using the vertical indicator ($76 \pm 9$ per cent).

Based on the absolute ages for 47 Tuc discussed earlier, we find that
from our horizontal method calculations, the age of ESO 121-SC03 lies in the range
$8.2 - 9.6 \pm 0.9$ Gyr.

\section{Discussion and Conclusions}
\label{s:conclusions}
In this paper we have presented photometric measurements from ACS/WFC imaging of 
ESO 121-SC03 in the F555W and F814W passbands. The resulting CMD represents the deepest 
and most precise published photometry for this unique LMC globular cluster. We have 
also presented new ACS/WFC photometry for the accreted Sagittarius dSph globular 
cluster Palomar 12. Using these data, we have conducted a detailed study
of the age of ESO 121-SC03. Specifically, we have derived its
age relative to the comparison clusters Pal. 12 and 47 Tuc using vertical and horizontal
dating methods calibrated against the stellar evolution models of \citet{vandenberg}.
These are well sampled in metallicity and allow us to account for the uncertain
run of $\alpha$-elements in ESO 121-SC03, as well as the fact that Pal. 12 is deficient
in $\alpha$-elements compared to many Galactic globulars, including 47 Tuc.

Our main result is that ESO 121-SC03 is significantly younger than 47 Tuc, but rather
similar in age to Pal. 12. This conclusion is independent of assumed $\alpha$-element 
abundance and age-dating method. Taking a straight error-weighted mean of our age 
measurements yields ESO 121-SC03 to be $73 \pm 4$ per cent the age of 47 Tuc, and 
$91 \pm 5$ per cent the age of Palomar 12. Palomar 12 is in turn $79 \pm 6$ per cent 
as old as 47 Tuc, in good agreement with previous measurements.
As discussed in Section \ref{ss:vertical}, recent results suggest the oldest Galactic 
globular clusters have ages of $\simeq 13.4$ Gyr. However, there is some disagreement 
in the literature as to whether 47 Tuc is coeval with the oldest Galactic globulars, or 
$\sim 15$ per cent younger. This uncertainty translates to an absolute age range for
ESO 121-SC03, spanning $8.3 - 9.8$ Gyr. Similarly, the absolute age of Pal. 12 lies in
the range $9.0 - 10.6$ Gyr. Typical errors on all these ages are $\pm 1.0$ Gyr.

Our new, deep photometry 
and precise age estimate fully confirms ESO 121-SC03 as the only known LMC cluster with 
an age placing it squarely within the age-gap. In this respect, it is clearly a unique 
object which may well be able to tell us important information about the formation and 
evolution of the LMC. Does this cluster represent the tail-end of ancient globular cluster 
formation in the LMC, just as Pal. 12 and Terzan 7 do for the Sagittarius dSph galaxy? 
If so, then how can we reconcile this picture and the observed metallicity of ESO 121-SC03
with its remote location in the very outer LMC halo? Other ancient globular clusters
at similar radii (such as NGC 1841 and Reticulum) are considerably more metal poor.

Prompted by the observation that ESO 121-SC03 is surrounded by a field population
which shares the same properties as the cluster, \citet{bica} speculated that
ESO 121-SC03 may represent part of a recently accreted LMC building block (or dwarf 
galaxy). We note however, that \citet{dirsch} found the age and metallicity of
this cluster to be consistent with the age-metallicity relation they derived from LMC
field star observations; hence they concluded the accretion of ESO 121-SC03 not a 
necessity. Even so, it would clearly be of interest to examine in more detail 
the possibility that ESO 121-SC03 is a captured object. 

By analogue with known accreted clusters in the Galactic halo (such as Pal. 12), we 
may expect three characteristics of ESO 121-SC03 if it is indeed a captured cluster.
First it may exhibit a peculiar velocity, at odds with that expected from the observed
dynamics of the LMC. Second, we may hope to observe the presence of some stream or
anomalous stellar population surrounding the cluster. Finally, it may be possible that
ESO 121-SC03 possesses elemental abundance patterns different to those observed in 
other LMC globular clusters -- just as Pal. 12 has been demonstrated to possess abundance
ratios unlike Galactic globular clusters or field stars, but in agreement with those
observed for the Sagittarius dSph \citep{cohen}. 

Regarding the first characteristic, we note that \citet{olszewski} measured a radial 
velocity of $309$ km$\,$s$^{-1}$ for ESO 121-SC03. At a position angle (measured
north through east) of $\Phi = 31\degr$ \citep{schommer} and angular
separation of $9.9\degr$ from the LMC centre, this velocity fits very well with the 
observed carbon star kinematics and circularly-rotating disk model summarized by 
\citet{marel} in their Figure 5. The observed velocity of ESO 121-SC03 also fits well 
with the rotational velocities exhibited by other old outer LMC globular clusters 
\citep[see][ Figure 6d]{schommer}. These results suggest that ESO 121-SC03 may not be
an accreted cluster, unless its parent building-block shared the kinematics characteristic
of the outer regions of the LMC.

Even so, regarding the second and third points above, it would clearly be of significant
interest to target the area surrounding ESO 121-SC03 with deep wide-field imaging, in
order to better survey the field star populations in this region. In would also be of
great value to obtain high resolution spectral measurements with the aim of determining
abundance ratios in this cluster. Together such observations may offer important clues
to the origin of ESO 121-SC03 with strong implications for the formation and evolution 
of the LMC.

\section*{Acknowledgements}
This paper is based on observations made with the NASA/ESA Hubble Space 
Telescope, obtained at the Space Telescope Science Institute, which is 
operated by the Association of Universities for Research in Astronomy, Inc., 
under NASA contract NAS 5-26555. These observations are associated with program
\#9891. ADM is grateful for financial support from a PPARC Postdoctoral Fellowship.
Many thanks also to Francesca de Angeli for useful discussions about blue stragglers,
and the anonymous referee whose comments helped improve the paper.

\bsp
\label{lastpage}

\end{document}